\documentclass[a4paper,12pt]{article}
\usepackage{graphicx, rotating}
\usepackage{ifpdf}
\ifpdf
\usepackage{hyperref, pdfsync, epstopdf, enumerate}	
\else
\usepackage[dvips,bookmarks]{hyperref}	
\fi
\hypersetup{colorlinks,bookmarksopen,bookmarksnumbered,citecolor=verdes,
linkcolor=blus,pdfstartview=FitH,urlcolor=rossos}
\def\hhref#1{\href{http://arxiv.org/abs/#1}{#1}} 

\renewcommand{\Re}{\,{\rm Re}}

\usepackage{amsfonts}
\usepackage{amsmath}
\usepackage{amssymb}
\usepackage{graphicx}%
\usepackage{slashed}

\newcommand{\beq}{\begin{equation}}
\newcommand{\eeq}{\end{equation}}
\newcommand{\fig}[1]{~\ref{fig:#1}}

\newcount\Mac  \Mac=1  
\newcommand{\ifMac}[2]{\ifnum\Mac=1 #1 \else #2 \fi}
\def\putps(#1,#2)(#3,#4)#5#6{\ifnum\Mac=1 \put(#1,#2){\special{picture #5}}
\else  \put(#3,#4){\includegraphics{#6}} \fi}

\newcommand{\One}{\hbox{1\kern-.24em I}}

\newcommand{\GeV}{\,{\rm GeV}}

\newcommand{\eV}{\,{\rm eV}}
\newcommand{\NP}{Nucl. Phys.}

\newcommand{\PL}{Phys. Lett.}
\newcommand{\PR}{Phys. Rev.}

\newcommand{\eq}[1]{~{\rm (\ref{eq:#1})}}

\newcommand{\pl}{p\hspace{-4.2pt}{\scriptstyle/}}

\newcommand{\lascia}[1]{}
\makeatletter
\def\art{\@ifnextchar[{\eart}{\oart}}
\def\eart[#1]#2#3#4#5#6{{\rm #2}, {#3 #4} {\rm (#6) #5} [{\hhref{#1}}]}
\def\hepart[#1]#2{{\rm #2, arXiv:\hhref{#1}}}
\newcommand{\oart}[5]{{\rm #1}, {#2 #3} {\rm (#5) #4}}

\newcounter{alphaequation}[equation]
\def\thealphaequation{\theequation\hbox to
0.6em{\hfil\alph{alphaequation}\hfil}}
\def\eqnsystem#1{
\def\@eqnnum{{\rm (\thealphaequation)}}
\def\@@eqncr{\let\@tempa\relax \ifcase\@eqcnt \def\@tempa{& & &} \or
  \def\@tempa{& &}\or \def\@tempa{&}\fi\@tempa
  \if@eqnsw\@eqnnum\refstepcounter{alphaequation}\fi
\global\@eqnswtrue\global\@eqcnt=0\cr}
\refstepcounter{equation} \let\@currentlabel\theequation \def\@tempb{#1}
\ifx\@tempb\empty\else\label{#1}\fi
\refstepcounter{alphaequation}
\let\@currentlabel\thealphaequation
\global\@eqnswtrue\global\@eqcnt=0 \tabskip\@centering\let\\=\@eqncr
$$\halign to \displaywidth\bgroup \@eqnsel\hskip\@centering
$\displaystyle\tabskip\z@{##}$&\global\@eqcnt\@ne
\hskip2\arraycolsep\hfil${##}$\hfil& \global\@eqcnt\tw@\hskip2\arraycolsep
$\displaystyle\tabskip\z@{##}$\hfil
\tabskip\@centering&\llap{##}\tabskip\z@\cr}

\def\endeqnsystem{\@@eqncr\egroup$$\global\@ignoretrue} \makeatother

\def\Tr{\mathop{\rm Tr}}
\def\circa#1{\,\raise.3ex\hbox{$#1$\kern-.75em\lower1ex\hbox{$\sim$}}\,}

\usepackage{multicol}
\usepackage{color}
\definecolor{rosso}{cmyk}{0,1,1,0.4}
\definecolor{rossos}{cmyk}{0,1,1,0.55}
\definecolor{rossoc}{cmyk}{0,1,1,0.2}
\definecolor{blu}{cmyk}{1,1,0,0.3}
\definecolor{blus}{cmyk}{1,1,0,0.6}
\definecolor{bluc}{cmyk}{1,1,0,0.1}
\definecolor{verde}{cmyk}{0.92,0,0.59,0.25}
\definecolor{verdec}{cmyk}{0.92,0,0.59,0.15}
\definecolor{verdes}{cmyk}{0.92,0,0.59,0.4}
\definecolor{grigio}{cmyk}{0,0,0,0.07}
\definecolor{rosa}{cmyk}{0,0.1,0.1,0.02}
\definecolor{rosino}{cmyk}{0,0.05,0.05,0.02}
\definecolor{rosas}{cmyk}{0,0.3,0.25,0.05}
\definecolor{celeste}{cmyk}{0.1,0,0,0.02}
\definecolor{giallino}{cmyk}{0,0,0.4,0.02}
\definecolor{rosso}{cmyk}{0,1,1,0.4}
\definecolor{rossos}{cmyk}{0,1,1,0.55}
\definecolor{rossoc}{cmyk}{0,1,1,0.2}
\definecolor{blu}{cmyk}{1,1,0,0.3}
\definecolor{bluc}{cmyk}{1,1,0,0.1}
\definecolor{blucc}{cmyk}{0.7,0.5,0,0}
\definecolor{viola}{cmyk}{0,1,0,0.6}
\definecolor{viola2}{cmyk}{0,1,0.2,0.6}
\definecolor{verde}{cmyk}{0.92,0,0.59,0.25}
\definecolor{verdec}{cmyk}{0.92,0,0.59,0.15}
\definecolor{verdes}{cmyk}{0.92,0,0.59,0.4}
\definecolor{verdino}{cmyk}{0.12,0,0.09,0.05}
\definecolor{giallo}{cmyk}{0,0,1,0}
\definecolor{gialloverde}{cmyk}{0.44,0,0.74,0}

\font\tenrsfs=rsfs10 at 12pt
\font\sevenrsfs=rsfs7
\font\fiversfs=rsfs5
\newfam\rsfsfam
\textfont\rsfsfam=\tenrsfs
\scriptfont\rsfsfam=\sevenrsfs
\scriptscriptfont\rsfsfam=\fiversfs
\def\mathscr#1{{\fam\rsfsfam\relax#1}}

\begin{document}
\color{black}
\vspace{1.0cm}

\begin{center}
\centerline{IFUP-TH 2008/19}\medskip
{\Huge\bf\color{rossos}Towards leptogenesis at NLO: the\\[3mm]
right-handed neutrino interaction rate}\\
\medskip
\bigskip\color{black}\vspace{0.6cm}
%
{\large\bf Alberto Salvio}$^a$,
{\large\bf Paolo Lodone}$^a$,
{\large\bf Alessandro Strumia}$^{b,c}$\\[4mm]
{\it (a) Scuola Normale Superiore di Pisa and INFN, Italia}\\[3mm]
{\it (b) Dipartimento di Fisica dell'Universit{\`a} di Pisa and INFN, Italia}\\[3mm]
{\it (c) NICPB, Ravala 10, 10143 Tallinn, Estonia}\\

\bigskip\bigskip

\centerline{\large\bf Abstract}
\begin{quote}\large
We compute quantum and thermal corrections to the right-handed neutrino
interaction rate in the early universe at next-to-leading order in all the relevant SM couplings
(gauge, top Yukawa  and higgs couplings).
Previous computations considered $2\to 2$ scatterings, 
finding infra-red divergences.
The KLN theorem demands that infra-red divergences cancel in the full result:
after adding  $1\to 3$ and one-loop virtual corrections that enter at the same order
we find a simple result.
\end{quote}
\end{center}

\vspace{-0.5cm}

{\small 
\tableofcontents
}

\newpage

\bigskip

\section{Introduction}
Thermal leptogenesis~\cite{fuk}  seems the most plausible explanation of
the observed baryon asymmetry of the universe~\cite{sak}.
A key quantity for leptogenesis is
the (space-time density of the) rate at which the thermal plasma  of the early universe at temperature $T$ creates quanta of
the lightest right-handed neutrino $N$ with mass $M$:
\beq \gamma_N(T) = \frac{dN_N}{dV\,dt}.\eeq
In thermal equilibrium, the creation rate equals the destruction rate, such that both quantities are usually named
``equilibrium interaction rate''.  It enters in the Boltzmann equation for the evolution of the total $N$ abundance $n_N$:
\beq \label{eq:Bol}sHz \frac{dY_{N}}{dz}= -\bigg(\frac{Y_{N}}{Y_{N}^{\rm eq}}-1\bigg)\gamma_N,\eeq
where $Y_N=n_N/s$, $s$ is the entropy density,
$z=M/T$ and $H(z)$ is the expansion rate.

\medskip

At leading order, $\gamma_N$ is given by the thermal average of
the $N\to LH, \bar L\bar H$ decay rate $\Gamma_N(E)$
induced at tree level by the Yukawa coupling $\lambda~NLH$, where $L$ and $H$ are the usual lepton and Higgs doublets
and $E=\sqrt{p^2+M^2}$  is the energy of $N$~\cite{k-sm,Plum1,GNRRS,Hannestad}:
\beq \label{eq:gamma0}
\gamma_N^{\rm LO}=2 \int \frac{d^3p}{(2\pi)^3} f_N \Gamma_N^{\rm LO}(E),\qquad
\Gamma_N^{\rm LO}(E) = \lambda^2 \frac{M}{8\pi} \frac{M}{E},\eeq
where $f_N = 1/(e^{E/T}+1)$ is the Fermi-Dirac distribution.

\medskip

{\em The goal of this paper is computing all quantum and thermal corrections to the $N$ interaction rate $\gamma_N$,
up to NLO in all relevant SM couplings $g$: the gauge couplings $g_2,g_Y$, the top Yukawa  coupling $\lambda_t=m_t/v$
and the Higgs quartic interaction $\lambda_h=(m_h/2v)^2$, where $v=174\GeV$ and $m_h$ is the zero temperature Higgs mass.}

\medskip

Previous partial results are extremely complicated because only some NLO effects have been computed,
missing the great simplification that happens when including all NLO corrections: 
infra-red (IR)  divergences cancel out in the total result, as demanded by the Kinoshita-Lee-Nauenberg (KLN) theorem~\cite{KNL}.
The final result must have the form

\beq\label{eq:gammaT} \gamma_N  = \gamma_N^{\rm LO}  \left[ 1+ K_0 \frac{g^2}{(4\pi)^2} +  K_T g^2 \frac{T^2}{M^2}+
{\cal O}(\frac{T}{M})^4 + \hbox{NNLO orders}\right],\eeq
where $K_0$ and $K_T$ are  order-one constants, computed in the rest of this paper.
\begin{itemize}
\item $K_0$ is the zero temperature quantum correction, that was so far ignored.

\item $K_T$ is the finite temperature correction, that is precisely needed only at $T\ll M$.
Indeed thermal leptogenesis does not depend on the initial $N$ abundance only if
$\lambda$ is  large enough that
right-handed neutrinos remain close to thermal equilibrium down to  low temperatures $T \ll M$;
we therefore only need a precision computation of $\gamma_N$ in such limit.
\end{itemize}
Some thermal effects have been computed so far~\cite{Plum1,GNRRS,BCST,Pilaftsis,Hannestad}: the contribution coming from
$2\to 2$ scatterings (such as $AN \to LH$,
where $A$ is any SM vector),
finding lengthy expressions where thermal masses regulate infra-red (IR) divergences~\cite{BCST,GNRRS}.
In thermal field theory this effect is just one correction to $\gamma_N$;
after adding all other effects (3-body decays, such as $N\to LHA$, and virtual corrections) we will find that IR divergences
cancel out.

In section~\ref{T=0} we compute quantum corrections.
In section~\ref{T} we compute thermal corrections.
The computations are lengthy, but thanks to cancellation of IR divergences the final result can be written in one line:
it is presented in the conclusions, section~\ref{concl}.

\section{Quantum corrections}\label{T=0}
In this section we compute the quantum correction to the $N$ interaction rate $\Gamma_N$,
up to  ${\cal O}(g_2^2, g_Y^2, \lambda_t^2, \lambda_h)$.  Such corrections have been neglected so far, and include two effects:
i) one loop corrections to $N\to LH$ and  ii) 3-body decays, such as $N\to LHA$.
Separately they are infra-red divergent. However,
as computed in the rest of this section and as demanded by the KLN theorem~\cite{KNL},
they combine to produce a $N$ decay rate that does not depend on the IR structure of the theory.

\subsection{Quantum corrections: tools}
We employ dimensional regularization for both IR and UV divergences.
The phase space in $d=4-2\varepsilon$ dimensions for one particle with quadri-momentum $P=(E,p)$
decaying into $n$ particles with quadri-momenta $P_i$ is:
\beq d\Phi_n = (2\pi)^d \delta(P-\sum_{i=1}^n P_i) \prod_{i=1}^n d\vec{p}_i,\qquad
d\vec{p}_i \equiv \frac{d^dP_i}{(2\pi)^d} {2\pi}\, \delta(P_i^2-m_i^2)=\frac{d^{d-1}p_i}{(2\pi)^{d-1}2E_i}.\eeq
For massless final-state particles the 2-body  phase space is 
\beq \Phi_2 = \frac{M^{d-4}}{2^{d-1} \pi^{d/2-1}}\frac{\Gamma(d/2-1)}{\Gamma(d-2)}\stackrel{d\to 4}{=} \frac{1}{8\pi} \eeq
and the 3-body phase space is
\beq \label{eq:dPhi3x1x2}
d\Phi_3 = \frac{M^2 e^{\gamma_E(4-d)}}{16(2\pi)^3} \left(\frac{M^2}{\bar\mu^2}\right)^{d-4} \frac{\left[
(1-x_1)(1-x_2)(1-x_3)\right]^{d/2-2}}{\Gamma(d-2)} dx_1 ~dx_2 \stackrel{d\to 4}{=}\frac{M^2~dx_1 dx_2}{128\pi^3},
\eeq
where $x_i \equiv 2 P_i\cdot P/P^2$ such that $x_1+x_2+x_3 =2$.
The integration region is $0<x_1<1$ and $1-x_1 <x_2 <1$: it is obtained
considering the triangle with sides $x_1,x_2,x_3$ and demanding that any side
is longer that the difference of the other two and shorter than their sum.

\medskip


We now compute the relevant corrections, in increasing order of difficulty: higgs, top and gauge.

\subsection{Higgs quantum correction}
There are no corrections induced at one loop by the quartic higgs coupling $\lambda_h$.

\subsection{Top quantum correction}\label{Top Yukawa T=0}
We compute the quantum corrections induced by the top quark Yukawa coupling $\lambda_t\, HQU$.
The only virtual correction is the correction to the $H$ propagator. All particles in the loop are massless, such that this correction
vanishes in dimensional regularization.  The only NLO correction is then the 3-body decay
$N\to LQU$.
We find:
\beq\Gamma(N\to LQU)=\Gamma_0 \frac{\lambda_t^2}{(4\pi)^2} \left(-\frac{3}{\varepsilon} + 3 \ell - \frac{21}{2}\right),\eeq
where here and in the following $\ell\equiv \ln M^2/\bar\mu^2$, with $\bar\mu$ being the 
$\overline{\rm MS}$ renormalization scale.
The UV divergence gets reabsorbed by writing the top Yukawa coupling in terms of its value
renormalized in the $\overline{\rm MS}$ scheme and the scale $\bar\mu=M$, obtaining, for the top quantum corrections at NLO:
\beq \Gamma_{N}^{{\rm top},T=0} = \frac{\lambda^2(M)M}{8\pi}\left[1- \frac{21}{2}\frac{\lambda_t^2}{(4\pi)^2}\right]. \label{topT=0}\eeq
Since IR divergences cancel, the same result can be obtained with different IR regularizations.\footnote{
Alternatively, one can regularize the IR divergence with a small top mass, $m_t \ll m_N$.
In such a case we get
\beq \Gamma(N\to LQU) = \Gamma_0 \frac{\lambda_t^2}{(4\pi)^2} \left( - \frac{23}{2} - 6\ln \frac{m_t}{M}\right) .\eeq
The virtual correction is both UV and IR divergent:
\beq \Gamma_{\rm virtual}(N\to LH) = \Gamma_0 \frac{\lambda_t^2}{(4\pi)^2} \left( -\frac{3}{\varepsilon} -3\ln\frac{\bar\mu^2}{m_t^2}+1\right) \eeq
giving again the same final result for $\Gamma_{N}^{{\rm top},T=0} = \Gamma_0 + \Gamma(N\to LQU) + \Gamma_{\rm virtual}(N\to LH)$.}

\subsection{Gauge quantum corrections}\label{gaugeT=0}
We consider one abelian vector with coupling $\alpha$ under which $L$ and $H$ have charge one; it will be easy to 
add at the end the group factors appropriate for SM vectors.
Virtual corrections to on-shell propagators of massless particles vanish in dimensional regularization.
Only the vertex diagram contributes to virtual corrections, and the result is:
\beq
\Gamma_{\rm virtual}(N\to LH) =\Gamma_0 \frac{\alpha }{4\pi} \left(-\frac{4}{\varepsilon^2} +4\frac{\ell-1}{\varepsilon} +2 \ell (2-\ell) + \frac{7\pi^2}{3}-8\right).\eeq
Emission of one vector $A$ from either the fermion $L$ or the scalar $H$ gives, in Feynman gauge:
\beq \label{eq:3bodyT=0}
\Gamma(N\to LHA) = \Gamma_0 \frac{\alpha}{4\pi}\left(\frac{4}{\varepsilon^2} + \frac{7-4\ell}{\varepsilon}+\ell(2\ell-7) - \frac{7\pi^2}{3} + \frac{45}{4}\right).\eeq
Summing real and virtual corrections the IR divergence cancels, leaving 
\beq 
\Gamma_{\rm NLO} =\Gamma_0\bigg[1 + \frac{\alpha}{4\pi}\bigg(\frac{3}{\varepsilon}+ 3\ln\frac{\bar\mu^2}{M^2}+\frac{29}{2}
\bigg)\bigg] = \frac{M \lambda^2(M)}{8\pi}\bigg[1 +\frac{29}{2}   \frac{\alpha}{4\pi}\ \bigg]
\eeq
having expressed $\Gamma_0$ in terms of $\lambda$ renormalized at $\bar\mu=M$ in the $\overline{\rm MS}$ scheme.
The result is IR convergent as guaranteed by the KLN theorem. Finally inserting the group factors appropriate for the SM electroweak vectors, $\alpha\to (\alpha_Y + 3 \alpha_2)/4$,
 we find:
\beq
\Gamma_N^{{\rm gauge},T=0} = \frac{M \lambda^2(M)}{8\pi}
 \bigg[1 + \frac{29}{32\pi}(3 \alpha_2 + \alpha_Y)\bigg].
\eeq
Since IR divergences cancel, the same result can be obtained with different IR regularizations.\footnote{Using  a small vector mass $m_A$ as infrared regulator we find
\beq \Gamma(N\to LHA) = \Gamma_0 \times  \frac{\alpha }{12\pi}
(\frac{87}{2}-2\pi^2 + 42\ln r +24 \ln r^2),\qquad
r = \frac{m_A}{m_N}\ll 1.\eeq

The correction to the kinetic terms  of a massless fermion and of a massless scalar due to a loop of an abelian vector with mass $m_V$
in a generic $\xi$ gauge are
$$ \pl P_L \bigg[ 1 - \frac{\alpha }{4\pi} ( \xi(\frac{1}{\varepsilon} + \ln \frac{\bar\mu^2}{m_V^2} )+ \xi -\frac{3}{2})\bigg],\qquad
p^2 \bigg[1 - \frac{\alpha }{4\pi}\left( (\xi-3)(\frac{1}{\varepsilon} + \ln \frac{\bar\mu^2}{m_V^2}) + \xi -\frac{5}{2}\right)\bigg].$$
Adding also the vertex diagram, the total virtual correction is $\xi$-independent (such that RGE equations for
the couplings are gauge-independent):
$$
 \Gamma_{\rm virtual} = \Gamma_0\bigg[1 + \frac{\alpha }{4\pi}\bigg(\frac{3}{\varepsilon}+ 3\ln\frac{\bar\mu^2}{M^2}
-14 \ln r - 8 \ln^2 r+ \frac{2\pi^2}{3}\bigg)\bigg].
$$}

\subsection{Summary}
Including quantum corrections at NLO in all relevant SM couplings, the $N$ decay rate does not receive
any IR-divergent correction and is:
\beq
\Gamma_N^{T=0} = \frac{M \lambda^2(M)}{8\pi}
 \bigg[1 + \frac{29}{32\pi}(3 \alpha_2 + \alpha_Y)- \frac{21}{2}\frac{\lambda_t^2}{(4\pi)^2}\bigg],
\eeq
where $\lambda(M)$ is the neutrino Yukawa coupling renormalized at the $\overline{\rm MS}$ scale $\bar\mu=M$.
It satisfies the well known RGE equation:
$$(4\pi)^2 \frac{d\lambda(\bar\mu)}{d\ln \bar\mu} = -\lambda\bigg [\frac{3}{4} g_Y^2 + \frac{9}{4}g_2^2 - 3 \lambda_t^2\bigg ].$$
RGE equations at LO have been computed in~\cite{RGE}, but the 
connection between $M$ and $\lambda(M)$ with neutrino masses has not yet been computed at NLO.

In the next section we consider thermal corrections.

\section{Thermal corrections}\label{T}
As already discussed in the introduction, we are interested in the dominant thermal corrections in the
low temperature limit, $T\ll M$.

At tree level, thermal corrections are exponentially suppressed by the Boltzmann factor  $e^{-M/T}$,
as clear by the thermal function $f_N$ in\eq{gamma0}. Such corrections have been included in previous works~\cite{GNRRS}.

At loop level,  thermal corrections are only power suppressed
(e.g.\ because in processes such as $N\to LHA$ the vector energy can be comparable to the temperature even at $M\gg T$):
we want to compute the dominant corrections  proportional to $(T/M)^2$.
Despite the $(T/M)^2$ suppression, thermal corrections are relevant because not suppressed by any $1/(4\pi)^2$ loop factor,
as anticipated in eq.\eq{gammaT}, where the coefficient $K_T$ is expected to be of order one.
Inserting such interaction rate into Boltzmann equations, we
find the correction to the baryon asymmetry shown in fig.\fig{corr}, equal to
$-1.5\% K_T$ ($-3.5\%K_T$) for $\tilde{m}_1 = m_{\rm atm}$ ($m_{\rm sun}$),
so that  thermal corrections are expected to be comparable to  quantum corrections and need to be computed. 

We compare our approach with previous works.
\begin{itemize}
\item Ref.~\cite{GNRRS} performed a resummation of the leading thermal corrections in the high-temperature regime, $T \circa{>}M$, 
approximatively described by  thermal masses:
\begin{eqnsystem}{sys:SMm}
m_H^2&=& \bigg(\frac{3}{16} g^2_2 + \frac{1}{16}
g^2_Y + \frac{1}{4} \lambda^2_t + \frac12 \lambda_h\bigg)T^2,\label{thermal-mH}\label{eq:mHT} \\
m_L^2&=& \bigg(\frac{3}{32} g^2_2 + \frac{1}{32}g^2_Y\bigg)T^2 \label{thermal-mL}, 
\end{eqnsystem}
We are interested in the low-temperature regime, that is the relevant one in the regime $\tilde{m}_1 \gg 10^{-3}\eV$ where
leptogenesis is computable independently of the initial $N$ abundance.
The thermal masses\footnote{The factor of $2$ in front of $m_L^2$
arises because thermal masses are conventionally defined at zero momentum $p$ (energy of a particle at rest with respect to the plasma).
Thermal dispersion relations are not relativistic, and for fermions the ``thermal'' mass squared at $p\gg T$ is $2m_L^2$.} $m^2_H$ and $2m^2_L$ of $L$ and $H$ reduce the phase space for $N\to LH$  decays, providing one
contribution to $K_T = -2 m_H^2 / T^2 =-3g_2^2/8+\cdots \approx -0.2$.

\begin{figure}
\begin{center}
\includegraphics{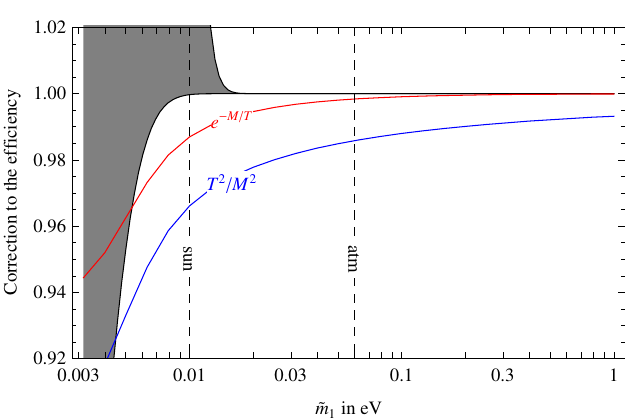}
\caption{\em Correction to the baryon asymmetry
due to thermal effects at tree level (red line, suppressed by $e^{-M/T}$) and at loop level
(blue line, suppressed by $T^2/M^2$).  The shaded region shows the uncertainty due to
the initial right-handed neutrino abundance, varied from negligible to dominant.
The dashed lines show the values  $\tilde{m}_1 = (\Delta m^2_{\rm sun})^{1/2}$ and 
 $(\Delta m^2_{\rm atm})^{1/2}$. \label{fig:corr}}
\end{center}
\end{figure}

\item  Another contribution to $K_T$ comes from $2\to 2$ scatterings, such as $AN\to LH$.
This process reduces to $N\to LH$ in the limit of vanishing energy of the vector $A$,
and it gives an IR divergent contribution to $K_T$ (it was computed in~\cite{GNRRS} using thermal masses as IR regulator).
Indeed such scattering rate is proportional to the number density of the initial vectors,
given by the Bose-Einstein distribution $n_A(\omega) = 1/[1 - e^{\omega/T}]$, that
diverges at small vector energy $\omega\to 0$, giving rise to the IR divergence.

\end{itemize}
These results are IR-divergent because incomplete: e.g.\ IR-divergent terms proportional to $n_A$ cancel
after adding $AN\to LH$ scatterings with $N\to LHA$ decays and with virtual corrections to $N\to LH$.

%


\bigskip





\subsection{Thermal corrections: tools}\label{tools}
\begin{figure}[t]
\begin{center}
\includegraphics[width=10cm]{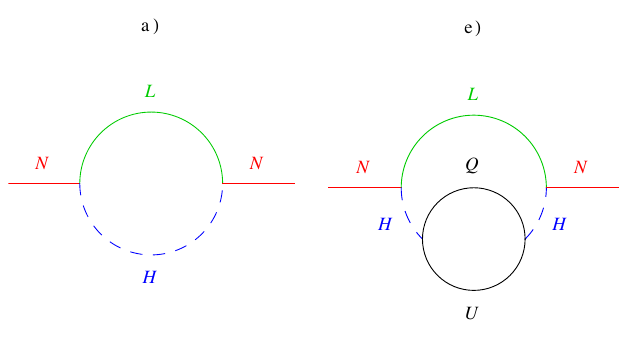}
\caption{\em The Feynman diagram for the top correction.\label{fig:Nprop}}
\end{center}
\end{figure}

We want to compute thermal corrections induced by the large
$g_Y, g_2,\lambda_t, \lambda_h$ couplings, while we can neglect those induced by the smaller neutrino Yukawa interaction $\lambda$.  In such a case the interaction rate at finite temperature is precisely defined and computed from
the imaginary part of
the $N$ propagator in the thermal plasma,
computed by explicitly summing all possible cuttings of the relevant
Feynman diagrams shown in fig.\fig{Nprop} and fig. \ref{Nprop-gauge}: the first one describes the tree-level result.

According to the real-time formalism of thermal field theory \cite{LeBellac}, the decay/absorption
rate $\Gamma$ of a particle with mass $M$ and quadri-momentum $P=(E,p,0,0)$
coupled to a thermal plasma by a weak coupling $\lambda$
is given by, at leading order in $\lambda$
\beq\label{eq:GammaT}
 \Gamma(E) = \frac{\Pi^>}{2E},
 \eeq
where $\Pi$ is the propagator (if the particle is a boson), or its spin-average (if the particle, as in our case, is a fermion:
$\Pi(P) = \Tr[ (\slashed{P}+M)\Sigma(P)]/2$
where $\Sigma$ is the fermion propagator).
$\Pi^>$ is the non time-ordered propagator, and can be 
computed following the rules of Kobes and Semenoff for imaginary parts of
Feynman diagrams at finite temperature, which generalize the cutting rules valid at $T=0$
for imaginary parts of Feynman diagrams.
$\Pi^>$ is essentially equivalent to ${\rm Im}\,\Pi$ and gives rise to simpler expressions.


The two-point function $\Pi^>(x_1,x_2)$ is computed summing over all possible ways of circling the internal type I $z_i$ vertices;
the external vertex $x_1$ is circled and the external  vertex $x_2$ is uncircled.
Non time-ordered correlations, such as $\Pi^>$, are computed using Feynman diagrams with the following additional rules:
\begin{itemize}
\item Reverse the sign of the coupling associated to a vertex, if it is circled.
\item For a propagator connecting two points $x$ and $y$:
\begin{enumerate}[i)]
\item Use the standard propagator $P(x-y)$ if neither $x$ nor $y$ are circled;
\item Use the propagator $P^*(x-y)$ if both $x$ and $y$ are circled;
\item Use the propagator $P^<(x-y)$ if $x$ but not $y$ is circled;
\item Use the propagator $P^>(x-y)$ if $y$ but not $x$ is circled;
\end{enumerate}
\end{itemize}
In momentum space, the propagators $P=\Delta_B$ for scalars, $P=(\slashed{K}+m)\Delta_F $ for fermions
(when $P^*$ is needed, the complex conjugate does not act on $\gamma_\mu$ matrices), 
$P=-g_{\mu\nu}\Delta_B$ for vectors in the Feynman gauge are
\beq\begin{array}{ll}
 \Delta_B = \Delta_0+2\pi n_B(K_0)\delta(K^2-m^2),\qquad&
\Delta_F = \Delta_0-2\pi n_F(K_0)\delta(K^2-m^2),\\[1ex]
\Delta^>_B =[\theta(+ K_0) + n_B(K_0)]  2\pi \delta(K^2-m^2),~&
\Delta^>_F=[\theta(+ K_0) - n_F(K_0)]  2\pi \delta(K^2-m^2),\\[1ex]
\Delta^<_B =[\theta(-K_0) + n_B(K_0)]  2\pi \delta(K^2-m^2),&
\Delta^<_F=[\theta(- K_0) - n_F(K_0)]  2\pi \delta(K^2-m^2),
\end{array}
\eeq
where $K$ is the quadri-momentum and $\Delta_0 ={i}/(K^2-m^2+i\epsilon)$ is the propagator at $T=0$;
$\Delta^>$  ($\Delta ^<$) are applied when $K$ enters into (exits from) the circled vertex.
The functions $n_{B,F}(E) =f_{B,F}(|E|)$ are the Bose-Einstein and
Fermi-Dirac statistical factors, 
$f_B(E)= 1/[e^{E/T}-1]$ and $f_F(E) = 1/[e^{E/T}+1]$.


After lengthy manipulations one finds that
ill-defined products of $\delta$ functions cancel out when summing over circlings of each type
of diagram separately, and one recovers a decomposition into `real' and `imaginary' contributions
and expressions similar to the ones well-known at $T=0$.


\bigskip

Thermal corrections can be written in a way similar to quantum corrections at $T=0$ by 
replacing propagators with thermal propagators and phase space with  `thermal phase space'.  
For a particle with mass $M$ and quadri-momentum $P$ this is defined as:
\beq \label{eq:dP}dP\equiv d\vec p\, \left[\theta(P_0) \pm n_{B,F}(P_0)\right] =
 \frac{d^dP}{(2\pi)^d}   \left[\theta(P_0) \pm n_{B,F}(P_0)\right]  2\pi  \delta(P^2-M^2),
 \eeq
which  generalizes the $T=0$ phase-space integral in $d=4-2\varepsilon$ dimensions.
Thermal processes such as the $N$ interaction rate $\Gamma_N$ combine contributions from
different quantum processes (e.g.\ decays such as $N\to LHA$ and scatterings such as $AN\to LH$): 
they are unified by the `thermal phase space' in\eq{dP} that allows each particle to have
positive $P_0$ (which means it is in the final state, as usual)
and negative $P_0$ (which means it is in the initial state).
Indeed the expression in square brackets gives the statistical factors appropriate for the two cases:
$1\pm n_{B,F}$ in the final state and $\pm n_{B,F}$ in the initial state. 
At $T=0$ the thermal factors $n_{B,F}$ vanish and the $\theta$-function forces $P_0=E>0$, recovering the 
usual phase space for final state particles: 
\beq \lim_{T\to 0} dP = d\vec p\equiv \frac{d^{d-1}p}{(2\pi)^{d-1}2E}.\eeq

\bigskip

For example, at leading order only the processes $N\to LH,\bar L\bar H$ are present
\beq \Gamma_N^{\rm tree}(E) = \frac{1}{2E}
\int dP_L~dP_H  \cdot (2\pi)^d \delta(P-P_H-P_L) \cdot\sum_{\rm final}|\mathscr{M}|^2, \eeq
where the sum is over polarizations and quantum numbers in all final states.
Kinematics demands that $L$ and $H$ can only be in the final state, such that the thermal average of such
decay rate is the usual expression~\cite{GNRRS}: 
\beq \label{gammaDBEFD}
 \gamma_N^{\rm tree} =  \int d\vec{p}_N d\vec{p}_L d\vec{p}_H f_N(1+f_H)(1-f_L)
(2\pi)^d \delta(P- P_L-P_H)\sum_{\rm all} |\mathscr{M}|^2,  \eeq
where the sum is over initial and final polarizations and quantum numbers.
Thermal effects $1\pm f$ break the usual Lorentz dilatation of the decay rate,
 $\Gamma(E)= \Gamma(M) M/E$, such that the integrals cannot be all done analytically.
 However these thermal effects are irrelevant at $T\ll M$, because suppressed by Boltzmann $e^{-{\cal O}(M)/T}$ factors,
and the tree result can be approximated  in terms of the decay width of a $N$ at rest at $T=0$:
\beq\label{eq:gammaGamma}
 \gamma_N^{\rm tree} \stackrel{T\ll M}{\simeq} n_N^{\rm eq}\,
 \Gamma_N(E=M)\times  \frac{{\rm K}_1(M/T)}{{\rm K}_2(M/T)}, \eeq
where the ratio of standard Bessel functions K$_1/{\rm K}_2\stackrel{T\ll M}{\simeq}1$ is the thermal average of 
$\Gamma(E)/\Gamma(M)$, approximated as $M/E$ according to the standard relativistic formula.

\bigskip

At NLO order in the gauge couplings one has an extra vector $A$ with quadri-momentum $K$ that enters in various
$3\to 1$ and $2\to 2$ processes, all described by the following expression
that resembles the $T=0$ result:
\beq \label{eq:Gammareal} \Gamma_N^{\rm real} = \frac{2}{2E}
\int dP_LdP_H dK  \cdot |\mathscr{M}|^2\cdot (2\pi)^d \delta(P-P_L-P_H-K),\eeq
where the phase-space factors $dP$ include the statistical factors for each particle, see eq.\eq{dP}.
The $L,H,A$ particles can be in the initial or in the final state, according to the sign of $P_L^0, P_H^0,K^0$:
thereby eq.\eq{Gammareal} unifies the $1\to 3$ decay with $2\to 2$ scatterings.

We see that we only need to compute at NLO the decay width $\Gamma_N$ of a $N$ at rest with respect to
the thermal plasma.

When computing thermal corrections we will also get (in the $T\to 0$ limit) the quantum corrections, already computed in section~\ref{T=0}.
They must be discarded keeping only purely thermal corrections, which are not affected by UV divergences.


\subsection{Higgs thermal correction}
The only thermal effect present at NLO is the higgs coupling contribution to the higgs thermal mass, see eq.\eq{mHT}.
{It reduces the decay width  for $N\to LH$ by a factor $1-2 m_H^2/M^2$, such that the higgs coefficient in eq.\eq{gammaT} is
\beq K_T^{\rm higgs} = -1. \eeq
%
}

\subsection{Top thermal correction}
We compute here the thermal corrections induced by the top quark Yukawa coupling $\lambda_t\, HQU$ at ${\cal O}(T^2/M^2)$.
We will find that the top coefficient in  eq.\eq{gammaT} is
{\beq K_T^{\rm top}= 0. \label{top-final-result}\eeq }
 Some of the formulae introduced in this section will be used to compute the gauge corrections in section \ref{gauge-thermal}.

%

\begin{figure}[t]
\begin{center}
\includegraphics[width=17cm]{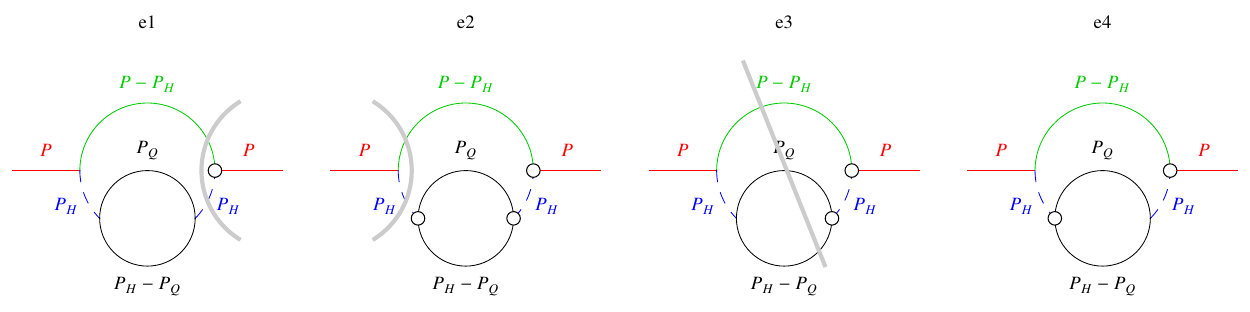}

\end{center}


\caption{\em Imaginary parts of the relevant Feynman diagrams for the top corrections, using the Kobes and Semenoff
circling notation~\cite{LeBellac}. When circled vertices form connected sets, we also show the corresponding
cutting notation that holds at $T=0$.
Momenta indicated in the diagrams flow from left to right.}\label{top-KS}
\end{figure}

\subsubsection{Virtual top thermal correction}\label{Virtual top thermal correction}
Let us first  consider the virtual top corrections, diagrams e1, e2 and e4 in fig.~\ref{top-KS}. The last diagram is proportional to an ill-defined $\delta$ function squared, 
that cancels exactly with the $\delta$ function squared coming from the other 
diagrams (including the real correction e3). We are therefore left with the remaining part of diagrams e1 and e2:
thermal self energy corrections to the $H$ line.  Their contribution can be written as:
\beq \Gamma^{\rm top}_{\rm self-energy}(E) =\Gamma_{\rm tree}(E) \left(Z_H^{\rm top}-1 -2 \frac{m_{H,{\rm top}}^2}{M^2}\right), \label{top-self-energy}\eeq
where $m_{H,{\rm top}}$ and $Z_H^{\rm top}$ are the top Yukawa contribution to the thermal mass and thermal wave function renormalization constant of $H$. These quantities 
can be computed from the top contribution to the thermal $H$ self-energy: 
\beq \Pi_t(P_H) \equiv \frac{3}{2} i \lambda_t^2 \mu^{4-d} \, {\rm Tr}\int \frac{d^d P_Q}{(2 \pi)^d} S(P_Q) S(P_Q-P_H),   \eeq
where $S(P_Q)$ is the thermal propagator of the quark, of which, as before, we neglect the mass. What actually appears in the calculation is not the full $H$ self-energy, but only its real part, ${\rm Re}\,\Pi_t$. Since Lorentz invariance is broken at finite temperature,
$\Pi_t$ does not only depend on $P_H^2$ but separately on $P_{H0}^2$ and $\vec{p}_{H}^{\, 2}$. The thermal mass  however 
turns out to be Lorentz invariant, and can be computed from ${\rm Re}\,\Pi_t(0)$, reproducing the standard result of eq.\eq{mHT}:
\beq m^2_{H,{\rm top}}= 6 \lambda_t^2 \int \frac{d^4P_Q}{(2\pi)^3}\delta(P_Q^2) n_F(P_{Q0}) = \frac{1}{4}\lambda_t^2T^2, \eeq
%
The wave function renormalization is given by the next to leading term in the expansion around $P_H^2=0$:
\beq Z_H^{\rm top} = 1 + \frac{\partial {\rm Re}\,\Pi_t}{\partial P_{H0}^2}\left(\vec{p}_{H}^{\,2},\vec{p}_{H}^{\,2}\right),\eeq
which generalizes the $T=0$ formula.  We obtain $Z_H^{\rm top}=1$. So the virtual top correction to the higgs decay width is 
just due to the reduction in phase space arising from the thermal Higgs mass~\cite{GNRRS}.  For $T\ll M$ we have:
\beq \Gamma^{\rm top}_{\rm self-energy}(E) =\Gamma_{\rm tree}(E) \left( -\frac{\lambda_t^2}{2}\frac{T^2}{M^2}\right)\qquad
\hbox{i.e.}\qquad
K_T^{\rm top, virtual} = -\frac{1}{2}
. \label{virtual-top-result}\eeq

\subsubsection{Real top thermal corrections}

The real correction  (diagram e3 in fig. \ref{top-KS}) leads to the decay/absorption rate as described in section~\ref{tools}
for the process  $N\to LQU,\bar L\bar Q\bar U$ with quadri-momenta $P=P_L+P_Q+P_U$:
\beq \label{eq:Gammareal-top} \Gamma^{\rm top}_{\rm real}(E) = \frac{2}{2E}
\int dP_L dP_QdP_U \cdot |\mathscr{M}_{\rm top}|^2\cdot (2\pi)^d \delta(P-P_L-P_Q-P_U),\eeq
where 
\beq |\mathscr{M}_{\rm top}|^2 =3 \lambda^2 \lambda_t^2 \frac{P\cdot  P_L}{P_Q \cdot  P_U} .  \label{Mtop2}\eeq
The 3-body phase space can be simplified as follows. We integrate over $dP_U$ using conservation of energy-momentum. Next:
\beq \delta(P_U^2) = \delta((P-P_L-P_Q)^2)=\delta(P^2-2P\cdot(P_L+P_Q)+2P_L\cdot P_Q). \label{delta-phase}\eeq
From now on, as previously discussed, we can assume that $P$ is at rest with respect to the plasma, $P=(M,0,0,0)$; then the statistical functions $n_{F}$ depend on $P_{L0}$, $P_{Q0}$ and $P_{U0}$ only and (\ref{delta-phase}) becomes
\beq \delta(M^2 - 2M(P_{L0}+P_{Q0})+2P_{L0}P_{Q0} -2|P_{L0}P_{Q0}|\cos\theta), \label{delta-phase-2}\eeq
where $P_{L0}$ and $P_{Q0}$ can be positive or negative
and $\theta$ is the angle between $\vec p_L$ and $\vec p_Q$.
The phase-space integral over $\theta$ is done using the $\delta$ function of eq.~(\ref{delta-phase-2}).
The condition $|\cos\theta|<1$ gives the allowed regions, that we now explicitly compute in terms of the two relevant free variables.

\begin{figure}
\begin{center}
$$\includegraphics[width=0.45\textwidth]{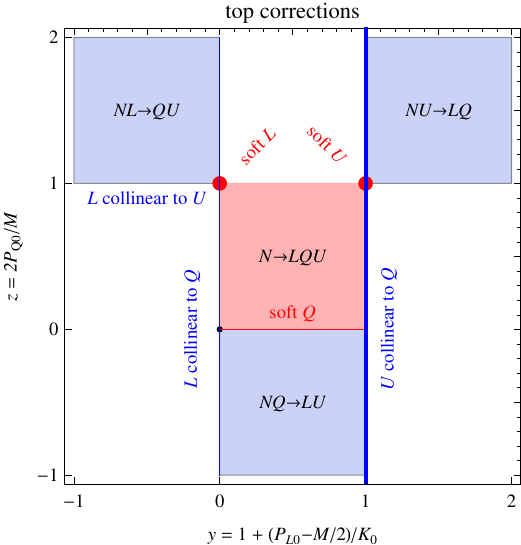}\qquad \includegraphics[width=0.45\textwidth]{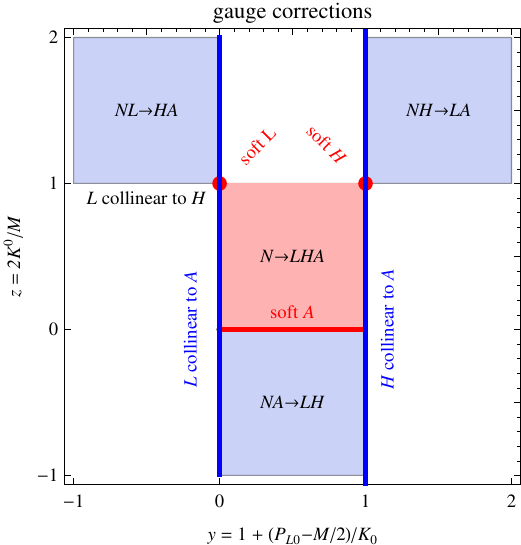}$$ \\

\hspace{0.7cm}{(a)} \hspace{7.7cm} {(b)}
\caption{\label{fig:regions}\em Kinematically allowed regions and their boundaries, where one has
soft and collinear singularities (thick lines) for top Yukawa (left) and gauge (right) thermal corrections.
Only the decay process is present at $T=0$.}
\end{center}
\end{figure}

Common choices are $P_{L0}$ and $P_{Q0}$, directly related to the dimensionless variables already employed in the quantum computation: $x_{L,U,Q}\equiv 2 P_{L,U,Q}\cdot P/P^2 = 2P_{L,U,Q0}/M$ ($x_L+ x_U + x_Q=2$).
It is more convenient to use two slighly different variables $z$ and $y$, defined by:
\beq
x_Q=z,\qquad
x_U=1-yz,\qquad
x_L=1-z(1-y)\eeq
i.e.\
\beq
P_{Q0} = \frac{M}{2} z,\qquad P_{U0} = \frac{M}{2}(1-yz),\qquad
P_{L0} = \frac{M}{2}(1+z(y-1)). \label{Pyz}
\eeq
Inverting these relations:
%
%
\beq z\equiv x_Q = 2\frac{P_Q\cdot P}{P^2},
\qquad
y \equiv 1+\frac{x_L-1}{x_Q}=1 + \frac{P_L\cdot P-P^2/2}{P_Q \cdot P}=
1+ \frac{P_{L0}-M/2}{P_{Q0}}.\qquad  \label{defyz}\eeq
The $z,y$ variables determine the angle as $|\cos\theta| =   |  1-2y/[1-(1-y)z]  |$ such that the condition $|\cos\theta|<1$
gives the kinematically allowed regions, plotted in fig.\fig{regions}a in terms of the
dimensionless variables $z,y$.
We just rediscover the well known allowed $1\to 3$ decay and $2\to 2$ scatterings.

We can now rewrite the  3-body phase space of\eq{dPhi3x1x2}, in terms of the $z,y$ variables:
\beq d\Phi_3 = 
\frac{M^2 e^{\gamma_E(4-d)}}{16(2\pi)^3} \left(\frac{M^2}{\bar\mu^2}\right)^{d-4} \frac{\left|y(1-y)(1-z)\right|^{d/2-2}}{\Gamma(d-2)} |z|^{d-3} dz~dy \stackrel{d\to 4}{=}\frac{M^2~|z|~dz~ dy}{128\pi^3}. \label{3body-yz}\eeq
 Also the amplitude of eq.~(\ref{Mtop2}) can be rewritten in terms of $y$ and $z$:
\beq |\mathscr{M}_{top}|^2= 3 \lambda^2 \lambda_t^2 \frac{1+(y-1)z}{(1-y)z}.  \eeq
By considering separately the various processes  corresponding to the four integration regions in the left plot of fig.~\ref{fig:regions},
we finally obtain the real top corrections as the sum of their contributions:
\begin{eqnsystem}{sys:topreal} 
\Gamma_{N\rightarrow L Q U} &=& \Gamma^0_{\rm tree} \frac{3 \lambda_t^2}{16 \pi ^2} \int_0^1 dy \int_0^1 dz (1-n_F(P_{L0}))(1-n_F(P_{Q 0}))(1-n_F(P_{U 0}))F(y,z), ~~\label{Gammareal1}\\
\Gamma_{N U \rightarrow L Q } &=& -\Gamma^0_{\rm tree} \frac{3 \lambda_t^2}{16 \pi ^2} \int_1^{\infty} dy \int_1^{\infty} dz (1-n_F(P_{L0}))(1-n_F(P_{Q 0}))n_F(P_{U 0})F(y,z), \label{Gammareal2}\\
\Gamma_{N Q\rightarrow L U } &=& -\Gamma^0_{\rm tree} \frac{3 \lambda_t^2}{16 \pi ^2} \int_0^1 dy \int_{-\infty}^0 dz (1-n_F(P_{L0}))n_F(P_{Q 0})(1-n_F(P_{U 0}))F(y,z), \label{Gammareal3}\ \\
\Gamma_{N L\rightarrow Q U } &=& -\Gamma^0_{\rm tree} \frac{3 \lambda_t^2}{16 \pi ^2} \int_{-\infty}^0 dy \int_1^{\infty} dz \, n_F(P_{L0})(1-n_F(P_{Q 0}))(1-n_F(P_{U 0}))F(y,z), \label{Gammareal4}
\end{eqnsystem}
where $\Gamma^0_{\rm tree}$ is the tree level decay rate at $T=0$ and 
\beq F(y,z)\equiv \frac{\left( e^{-\gamma_E }\frac{M^2}{\bar\mu^2}\right)^{d/2-2}}{\Gamma(d/2-1)} \frac{\left| y (1-y) (1-z)\right|^{d/2-2}[1+(y-1)z]}{(1-y)z|z|^{3-d}}.
  \eeq
We can now check that infrared divergences cancel after summing the different processes.
The situation is relatively simple:
\begin{itemize}
\item There are no thermal IR singularities, because we only have fermionic distributions, which are not singular at $n_F(0)$;
\item There is no soft singularity at $z\to 0$: the infrared singularity in $|\mathscr{M}_{\rm top}|^2$ gets canceled by positive powers of $z$ in the 3-body phase space in (\ref{3body-yz});
\item There is a collinear divergence, described by the $1/(1-y)$ factor at $y\to 1$.
As clear from fig.\fig{regions} this corresponds to $Q$ collinear with $U$ and the first 3 processes
in eq.~(\ref{sys:topreal}) are separately infra-red divergent.
\end{itemize}
One infrared divergent process is the decay $N\to LQU$ which was not included in previous works, that thereby missed
the cancellation of infrared divergences, which takes place in the full result, as we now describe.

The quantities in (\ref{Gammareal1})-(\ref{Gammareal3}) have the following poles at $d=4$:
\begin{eqnsystem}{sys:topdiv} 
\Gamma^{\rm div}_{N\rightarrow L Q U} &=& \frac{1}{d/2-2}\frac{M \lambda^2}{8 \pi} \frac{3 \lambda_t^2}{16 \pi ^2}  [1-n_F(\frac{M}{2})]\int_0^1 dz [1-n_F(\frac{Mz}{2})]
[1-n_F(\frac{M(1-z)}{2})],  \qquad \label{Gammarealdiv1}\\
\Gamma^{\rm div}_{N U\rightarrow L Q } &=& \frac{1}{d/2-2}\frac{M \lambda^2}{8 \pi} \frac{3 \lambda_t^2}{16 \pi ^2}[1-n_F(\frac{M}{2})] \int_1^{\infty} dz  [1-n_F(\frac{Mz}{2})]n_F(\frac{M(1-z)}{2}), \label{Gammarealdiv2}\\
\Gamma^{\rm div}_{N Q\rightarrow L U } &=& \frac{1}{d/2-2}\frac{M \lambda^2}{8 \pi} \frac{3 \lambda_t^2}{16 \pi ^2}[1-n_F(\frac{M}{2})] \int^0_{-\infty} dz~  n_F(\frac{Mz}{2})[1-n_F(\frac{M(1-z)}{2})]. \label{Gammarealdiv3}
\end{eqnsystem}
The integrations over $z$ can be performed analytically, with the result: 
\beq\label{eq:topIRtot}
\Gamma^{\rm div}_{N\rightarrow L Q U} +\Gamma^{\rm div}_{N U\rightarrow L Q } +\Gamma^{\rm div}_{N Q\rightarrow L U } =
 \frac{1}{d/2-2}\frac{M \lambda^2}{8 \pi} \frac{3 \lambda_t^2}{16 \pi ^2}  [1-n_F(\frac{M}{2})][1+n_B(\frac{M}{2})].
\eeq
In the limit $T\to 0$ (i.e.\ neglecting thermal functions) we recover the 
infra-red divergent real quantum correction (already computed in section \ref{Top Yukawa T=0}), here contained in 
$\Gamma_{N\to LQU}^{\rm div}$.
For generic $T$,  integrals over fermionic thermal functions squared produce the bosonic thermal function $1+n_B(M/2)$,
such that the total real infra-red divergence in real processes has the same structure of the tree level result.

Such real IR divergence cancels out with the IR divergence in the virtual contribution.  
The virtual contribution vanishes because it has opposite IR and UV divergences, not distinguished by dimensional regularization.
So, formally, the cancellation takes place when expressing the tree-level result in terms of 
the renormalized coupling:
\beq \lambda^2=\lambda^2(M)\left(1-\frac{1}{d/2-2} \frac{3\lambda_t^2}{16 \pi^2}+...\right),  \label{lambdaMSbar}
 \eeq
where $\lambda(M)$ is, as before, the $\overline{\rm MS}$ coupling at the scale $\bar\mu=M$ and the dots are the contributions of the other relevant coupling constants.
 Let us see in detail how this occurs. The tree level contribution at finite temperature can be written as 
\beq \Gamma_{\rm tree}(M)= \frac{M \lambda^2}{8\pi} \left( \frac{M^2}{\bar\mu^2}\right)^{d/2-2} \frac{ e^{\gamma_E (4-d)/2}\Gamma(d/2-1)}{\Gamma(d-2)} [1-n_F(\frac{M}{2})][1+n_B(\frac{M}{2})].\label{GammatreeT}\eeq
Substituting (\ref{lambdaMSbar}) into (\ref{GammatreeT}) and setting $\bar\mu=M$ one obtains a divergent part
which is exactly opposite to\eq{topIRtot}.

\bigskip

Having verified that IR divergences cancel, the practical recipe to get the finite result is to sum the finite parts of the various contributions:
\beq  \Gamma_{N\rightarrow L Q U}-\Gamma^{\rm div}_{N\rightarrow L Q U}+\Gamma_{N U\rightarrow L Q }-\Gamma^{\rm div}_{N U\rightarrow L Q }
+\Gamma_{N Q\rightarrow L U }-\Gamma^{\rm div}_{N Q\rightarrow L U }+\Gamma_{N L\rightarrow Q U }. \label{real-top} \eeq
The integrals in (\ref{Gammareal1})-(\ref{Gammareal4}) can be computed numerically for arbitrary values of $T/M$
and analytically for small $T/M$, following the appendix of  \cite{Altherr:1989jc}. 
%

We can neglect all  terms exponentially suppressed by factors $\sim e^{-{M}/{T}}$.
Each $n_F(E)$ in (\ref{Gammareal1})-(\ref{Gammareal4}) provide such suppression factor,
unless kinematics allows integration regions where $E\ll T$.
Kinematics allows at most one particle among $Q,U,L$ to have energy $E \ll T$;
thereby we can keep only the terms of~(\ref{sys:topreal}) which have one thermal function
(dropping the $T=0$ term with no thermal functions and the terms with two thermal functions):
\begin{eqnsystem}{sys:toplowT} \Gamma_{N\rightarrow L Q U} &\simeq&- \Gamma^0_{\rm tree} \frac{3 \lambda_t^2}{16 \pi ^2} \int_0^1 dy \int_0^1 dz (n_F(P_{L0})+n_F(P_{Q 0})+n_F(P_{U 0}))F(y,z), \label{Gammareal1exp}\\
\Gamma_{N U\rightarrow L Q } &\simeq& -\Gamma^0_{\rm tree} \frac{3 \lambda_t^2}{16 \pi ^2} \int_1^{\infty} dy \int_1^{\infty} dz\, n_F(P_{U 0})F(y,z), \label{Gammareal2exp}\\
\Gamma_{N Q\rightarrow L U } &\simeq& -\Gamma^0_{\rm tree} \frac{3 \lambda_t^2}{16 \pi ^2} \int_0^1 dy \int_{-\infty}^0 dz\, n_F(P_{Q 0})F(y,z), \label{Gammareal3exp}\ \\
\Gamma_{N L\rightarrow Q U } &\simeq& -\Gamma^0_{\rm tree} \frac{3 \lambda_t^2}{16 \pi ^2} \int_{-\infty}^0 dy \int_1^{\infty} dz \, \, n_F(P_{L0})F(y,z) \, . \label{Gammareal4exp}
\end{eqnsystem}
Finally, we integrate along the phase space where some particle has $E\ll T$:
\begin{itemize}
\item {\it Region $P_{Q 0}\sim 0$},  corresponding to the line $z \sim 0$ (see fig.\fig{regions}a)
The relevant terms are the second one of (\ref{Gammareal1exp}) plus (\ref{Gammareal3exp}).
We expand $F(y,z)$ in powers of $z$. To this end notice
\beq (1 - z)^{d/2 - 2} [1 + (y - 1) z] =   1 + (1 - d/2 + y) z + {\cal O}(z^2). \label{exp1} \eeq
The ${\cal O}(z)^0$ part on the right hand side of (\ref{exp1}) gives no contribution to $K_T$: indeed its contribution to the second term in (\ref{Gammareal1exp})
is opposite to that to (\ref{Gammareal3exp}) up to higher orders in $(T^2/M^2)$. 
So the only contribution at first order in $T^2/M^2$ comes from the ${\cal O}(z)$ part and we obtain
\beq \Delta\Gamma (|P_{Q 0}|\sim 0) \simeq \Gamma^0_{\rm tree}\frac{\lambda_t^2}{4} \frac{T^2}{M^2}. \label{K00}\eeq
\item 

{\it Region $P_{U 0}\sim 0$}, corresponding to  the point $y\sim z\sim 1$ (see fig.\fig{regions}a). The relevant terms are the third one of (\ref{Gammareal1exp}) plus (\ref{Gammareal2exp}).
Let us first expand $F(y,z)$ in powers of $(z-1)$ by noticing
\beq \frac{1}{ z^{4 - d}}[1 + (y - 1) z]= y + (d y-1 - 3 y ) (z - 1) + {\cal O}((1-z)^2). \label{exp2} \eeq
The contribution coming from the ${\cal O}(1)$ part in (\ref{exp2}) vanishes up to ${\cal O}(T^2/M^2)$, while the ${\cal O}(1-z)$ part gives
\beq \Delta\Gamma (|P_{U 0}|\sim 0) \simeq \Gamma^0_{\rm tree}\frac{\lambda_t^2}{4} \frac{T^2}{M^2}.  \label{K010}\eeq

\item 
{\it Region $P_L^0\sim 0$}, corresponding to the point $y \sim 0$ and $z\sim 1$ (see fig.\fig{regions}a). The relevant terms are the first one of (\ref{Gammareal1exp}) plus (\ref{Gammareal4exp}).
We find none of these terms contribute up to ${\cal O}(T^2/M^2)$. 
\end{itemize}
If we now sum (\ref{K00}) with (\ref{K010}) we obtain $K_T^{\rm top, real} = 1/2$. 
So the real plus virtual corrections, eq. (\ref{virtual-top-result}), give the final result anticipated in (\ref{top-final-result}):  $K_T^{\rm top} = 0$.

\subsection{Gauge thermal correction} \label{gauge-thermal}

We now turn to the thermal gauge corrections to the interaction rate of $N$ up to ${\cal O}(T^2/M^2)$ or, in other words, 
the gauge corrections to $K_T$ in eq. (\ref{eq:gammaT}). Our result is
\beq K_T^{\rm gauge}=0.  \label{gauge-result}\eeq

\begin{figure}
\begin{center}
\includegraphics[width=\textwidth]{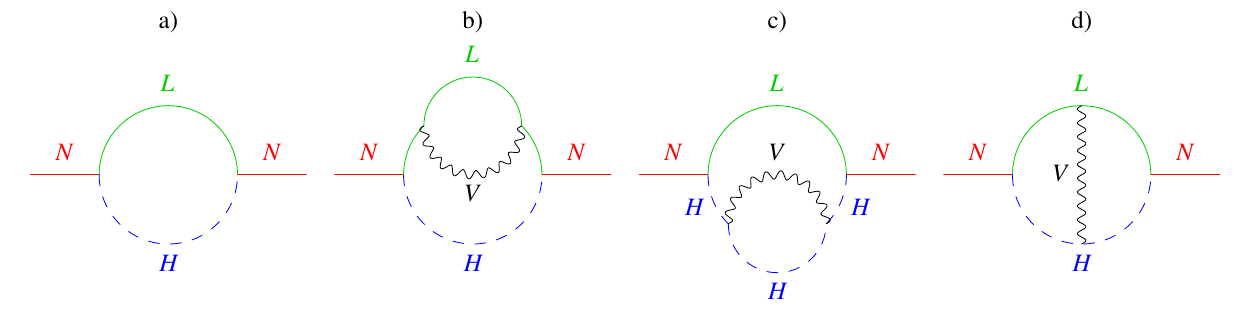}
\caption{\em The Feynman diagrams for gauge corrections.\label{Nprop-gauge}}
$$\includegraphics[width=\textwidth]{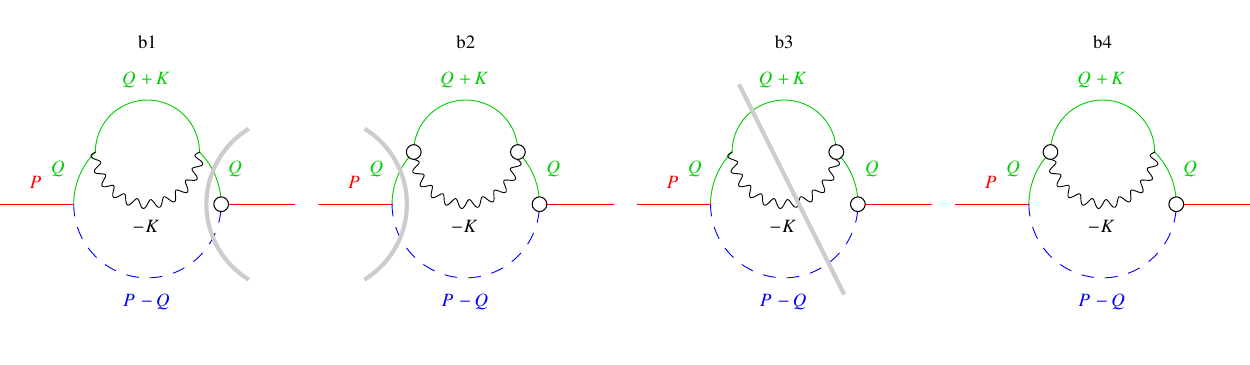}$$
$$\includegraphics[width=\textwidth]{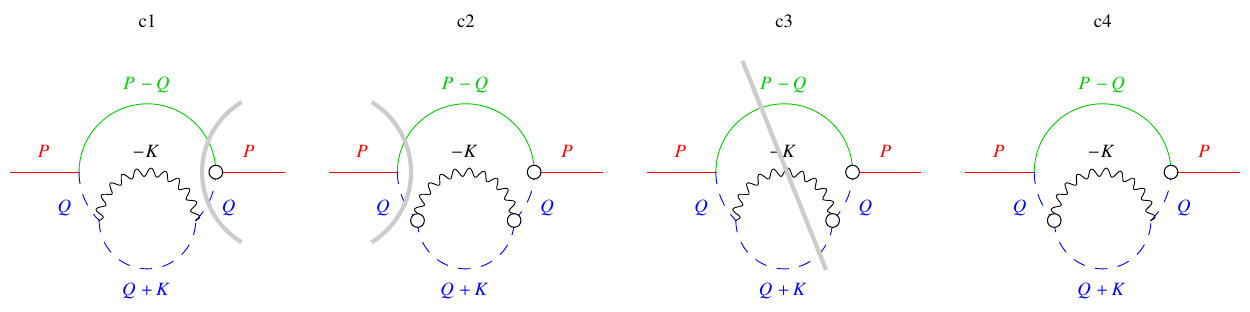}$$
$$\includegraphics[width=\textwidth]{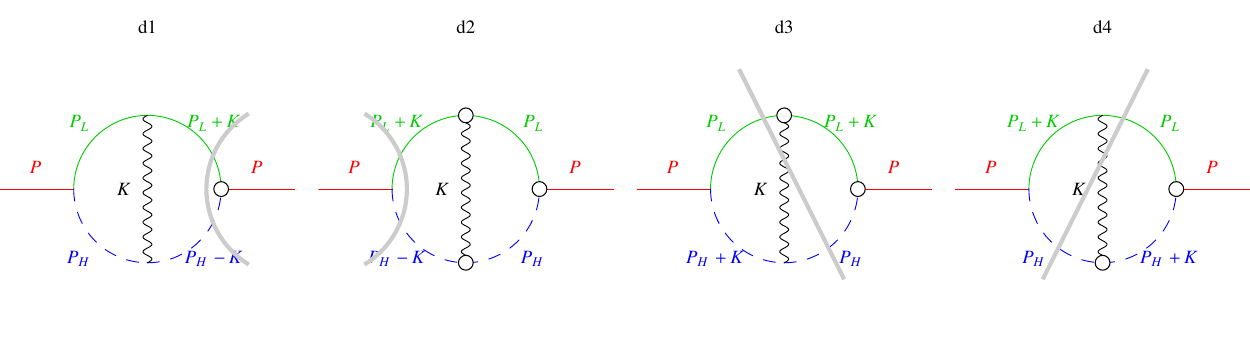}$$
\caption{\em Imaginary parts of the relevant Feynman diagrams for the gauge corrections. We use the same notation as in fig. \ref{top-KS}. }\label{fig:gauge-KS} 
\end{center}
\end{figure}

\subsubsection{Virtual gauge thermal corrections}

Let us start with the virtual corrections. We divide the contributions in vertex correction (diagrams d1 and d2 in Fig. \ref{fig:gauge-KS}), corrections to the $L$ propagator 
(diagrams b1, b2 and b4) and corrections to the $H$ propagator (c1, c2 and c4). These diagrams separately contain  ill-defined squares of $\delta$ functions, which, like in the top case, cancel when one sums them with 
the real corrections (diagrams b3, c3, d3 and d4).

\bigskip

We first consider the virtual contribution from the vertex diagrams d1 and d2. 
Like for the real part (see section \ref{tools}), the KS rules again give
an expression that resembles the $T=0$ result:  
\beq \Gamma_{\rm vertex}=-\frac{2}{2E}
\lambda^2   \times \mu^{8-2d}
\int dP_L ~dP_H~(2\pi)^d\delta(P-P_L-P_H)\times g^2\int \frac{d^dK}{(2\pi)^d} \mathscr{N}\,  \mathscr{P}, \label{vertex-integral}\eeq
where $g^2 \equiv 3 g_2^2/4+g_Y^2/4$;
the first integral is over the `thermal phase space' of the two-body decay;
the second integral is the loop integral over the vector quadri-momentum $K$;
$\mathscr{N}$ is the usual numerator coming from the Feynman diagram:
\beq \mathscr{N}= (2P_H-K)_\rho \Tr[ \slashed{P}\slashed{P}_L\gamma_\rho (\slashed{P}_L+\slashed{K})]\ ,\eeq
and $\mathscr{P}$ is the product of the various thermal propagators $\Delta_{B,F}$, that can be expanded in terms of zero
temperature propagators $\Delta_0$ as follows:  
\begin{eqnarray} \mathscr{P}&=&\Re[ \Delta_F (P_L+K) \Delta_B(P_H-K) \Delta_B(K)] = \nonumber \\
&=&\Re[ \Delta_0 (P_L+K) \Delta_0(P_H-K) \Delta_0(K)]+ \nonumber \\
&&+2\pi \delta(K^2)n_B(K_0) \Re [ \Delta_0 (P_L+K) \Delta_0(P_H-K) ]+  \label{eq:props}\\
&&+2\pi \delta((P_H-K)^2) n_B(P_H^0-K^0) \Re[\Delta_0(P_L+K)\Delta_0(K)]+\nonumber \\
&&-2\pi\delta((P_L+K)^2)n_F(P_L^0+K^0)\Re[\Delta_0(P_H-K)\Delta_0(K)].  \nonumber
\end{eqnarray}
All other terms with higher powers of $n_{B,F}$ vanish due to conflicting $\delta$-function requirements.
As usual we can assume $N$ at rest with respect to the plasma, $P=(M,0,0,0)$, and defining $c=\cos\theta$ as the angle between $\vec{p}_L$ and $\vec{k}$ we have
\beq \Delta_0 (P_L+K) =\frac{i}{2P_L\cdot K} = \frac{i}{M(K_0-|K_0 |c)},\qquad
\Delta_0 (P_H-K)=\frac{-i}{M(K_0 + |K_0| c)}.
\eeq
The loop integral of the first thermal term can be reduced to an integral over $c$ and $z\equiv 2|\vec k|/M>0$ as follows:
\beq \int \frac{d^dK}{(2\pi)^d}2\pi \delta(K^2) f(K_0, \vec k)= \frac{(M)^{d-2}}{\Gamma(d/2-1)(4\pi)^{d/2}} \int_0^\infty  dz~(z/2)^{d-3}\times\qquad\eeq
$$\times \int_{-1}^1\frac{dc}{(1-c^2)^{2-d/2}}
[f(Mz/2,\vec k)+f(-M z/2,\vec k)]. $$
The last two terms in\eq{props} can be similarly computed, by shifting the integration variable $K$ 
in order to obtain the same factor $\delta(K^2)n_{B,F}(K_0)$ for all three factors.
By dropping the $T=0$ contribution
in $\mathscr{N}\,  \mathscr{P}$, we obtain 
\begin{eqnarray}
 \Gamma_{\rm vertex}^{\rm thermal} (M)&\simeq& \Gamma_{\rm tree}^0
\frac{g^2}{16\pi^2}\frac{\left( e^{-\gamma_E }{M^2}/{\bar\mu^2}\right)^{d/2-2}}{\Gamma(d/2-1)} \int_0^\infty dz \int_{-1}^1 dc \frac{(z/2)^{d-3}}{(1-c^2)^{3-d/2}} \times \nonumber\\
&&\times
\left\{-\frac{32}{z^2}n_B(Mz/2)
-8\frac{1+c^2}{1-z^2} (n_B(Mz/2)-n_F(Mz/2))\right\}. \label{thermal-vertex}
\end{eqnarray}
This expression has collinear divergences at  $c=\pm 1$ (vector collinear with $H$ or $L$)
and soft divergences at $z=0$ (vector with vanishing energy).
There are no divergences at $z=1$; the apparent ones are  treated taking the principal part.
The structure of the singularities is plotted in fig. \ref{fig:regions}b.

\bigskip

Turning now to the self-energy contributions, we observe that they can be written as in (\ref{top-self-energy}) with $m^2_{H,\rm top}$ replaced with $m^2_{H,\rm gauge}+m_{L,\rm gauge}^2$ and $Z_H^{\rm top}-1$ with 
$Z_H^{\rm gauge}+Z_L-2$.
The definition of the $H$ thermal mass and wavefunction renormalization given in section \ref{Virtual top thermal correction} also holds in the gauge case (of course substituting $\Pi_t$ with the gauge contribution to the 
$H$ self-energy). We obtain $Z_H^{\rm gauge}=1$ and recover the known value of the thermal mass, $m^2_{H,\rm gauge}=g^2T^2/4$.
The calculation of the $L$ self-energy  requires more care. 
Like in the top case, what actually enters in the calculation is not the full $L$ self-energy,
but only its real part, which we call $\Sigma(P_L)$:
\begin{eqnarray*} \Sigma(P_L) &=& ig^2 \mu^{4-d}
\int\frac{d^dK}{(2\pi)^d} \Delta_B(K) \Delta_F(P_L+K)
\gamma_\mu (\slashed{P_L}+\slashed{K})\gamma_\mu\\
&=&2g^2 \mu^{4-d}(d/2-1)  \int\frac{d^dK}{(2\pi)^d}  2\pi \delta(K^2) \frac{(\slashed{K}+\slashed{P_L})n_B(K) + \slashed{K} n_F(K)}{(P_L+K)^2}\\
&=&g^2 \mu^{4-d}(d/2-1)  \int\frac{d^dK}{(2\pi)^{d-1}}  \delta(K^2) \frac{\slashed{K}}{P_L\cdot K}
[n_B(K_0)+n_F(K_0)], \qquad\hbox{at $P_L^2=0$}.
\end{eqnarray*}
The thermal mass can be defined again as the pole of the corrected propagator, $1/(\slashed{P}_L-\Sigma)$: squaring
 $(\slashed{P}_L - \Sigma)^2 \simeq P_L^2 - \left\{\slashed{P}_L,\Sigma\right\}$
we obtain the same expression as in eq.~(\ref{thermal-mL}): 
\beq m_{L,\rm gauge}^2 =\left\{\slashed{P}_L,\Sigma\right\}=2g^2 \int\frac{d^4K}{(2\pi)^3} \delta(K^2)[n_B(K_0) + n_F(K_0)]=
\frac{g^2}{4}T^2.\eeq
$Z_L$ is the residue at the pole, explicitly given by $ Z_L=1-\Tr( \gamma_0 \Sigma)/4P_{L0}$
in the rest frame of the plasma, such that  (see e.g.~\cite{Altherr:1988bg}):
\beq Z_L=1-\frac{4 g^2}{16 \pi^2} \left(\frac{M}{\bar \mu}\right)^{d/2-2}
\frac{d/2-1}{d/2-2}\int_0^{\infty}dz~z^{d-3}  (n_B(Mz/2)+n_F(Mz/2)). \label{ZL}
\eeq
$Z_L$ contains infra-red divergences, which emerge when the loop momentum $\vec{k}$ is parallel to $\vec{p}_L$ and are therefore of collinear type.

\subsubsection{Real gauge thermal corrections}

As already mentioned, the real corrections (diagrams b3, c3, d3 and d4 in Fig. \ref{fig:gauge-KS}) take the form (\ref{eq:Gammareal}). To simplify the phase space we can proceed 
like in the top case and repeat the steps in Eqs. (\ref{delta-phase}), (\ref{delta-phase-2}),
(\ref{defyz}), (\ref{Pyz}) and (\ref{3body-yz}) with the substitutions $P_Q\rightarrow K$ and $P_U \rightarrow P_H$.
 In the gauge case we obtain
\beq |\mathscr{M}_{\rm gauge}|^2 = 2\lambda ^2 g^2 \frac{2(1-z)+z^2(1-y)(d/2-1)}{z^2y(1-y)}.\eeq
Notice that $|\mathscr{M}_{\rm gauge}|^2$ has soft ($z=0$) and collinear ($y=\pm 1$) singularities. These lead to IR divergences in the real corrections as shown in  fig.~\ref{fig:regions}b.
 Like in the top case, we now express the real corrections as the sum of the following four integrals over $y$ and $z$, corresponding to the four regions in fig. \ref{fig:regions}:
\begin{eqnsystem}{sys:gaugereal}
 \Gamma_{N\rightarrow L H A} &=& \Gamma^0_{\rm tree} \frac{g^2}{16 \pi ^2} \int_0^1 dy \int_0^1 dz (1-n_F(P_{L0}))(1+n_B(K_0))(1+n_B(P^0_H))F_g(y,z),~~ \label{Gammarealg1}\\
\Gamma_{N H\rightarrow L A } &=& \Gamma^0_{\rm tree} \frac{g^2}{16 \pi ^2} \int_1^{\infty} dy \int_1^{\infty} dz (1-n_F(P_{L0}))(1+n_B(K_0))n_B(P^0_H)F_g(y,z), \label{Gammarealg2}\\
\Gamma_{N A\rightarrow L H } &=& \Gamma^0_{\rm tree} \frac{g^2}{16 \pi ^2} \int_0^1 dy \int_{-\infty}^0 dz (1-n_F(P_{L0}))n_B(K_0)(1+n_B(P^0_H))F_g(y,z), \label{Gammarealg3}\ \\
\Gamma_{N L\rightarrow H A} &=& -\Gamma^0_{\rm tree} \frac{g^2}{16 \pi ^2} \int_{-\infty}^0 dy \int_1^{\infty} dz \, n_F(P_{L0})(1+n_B(K_0))(1+n_B(P^0_H))F_g(y,z), \label{Gammarealg4}
\end{eqnsystem}
where  
\begin{eqnarray}
P_{L0} &=& \frac{M}{2} [1+z(y-1)], \quad K_0 =\frac{M}{2} z, \quad P^0_H= \frac{M}{2} (1-zy), \\
 F_g(y,z)&\equiv& 2 \frac{\left( e^{-\gamma_E }{M^2}/{\bar\mu^2}\right)^{d/2-2}}{\Gamma(d/2-1)} \frac{\left| y (1-y) (1-z)\right|^{d/2-2}[2(1-z)+(1-y)z^2(d/2-1)]}{y(1-y)z^2|z|^{3-d}}.
  \end{eqnarray}
Real corrections harbour collinear divergences from $y=0$ and $y=1$ and soft divergences from $z=0$. 
Apart from the usual infra-red divergences (arising from soft and collinear configurations where $|\mathscr{M}_{\rm gauge}|^2$ is singular), there are new purely thermal infra-red divergences, arising from configurations where
a boson has vanishing energy such that their bosonic thermal distribution $n_B$ diverge,
as anticipated at the beginning of section \ref{T}.

\subsubsection{Sum of the virtual and real contributions}
In principle, we should now proceed to compute explicitly the integrals in (\ref{thermal-vertex}), (\ref{ZL}) and~(\ref{sys:gaugereal}). 
In practice they are so difficult that also checking the cancellation of IR divergences would need ad hoc tricks.
In order to proceed in a systematic way, from now on we focus on the relevant limit $T\ll M$, dropping all exponentially suppressed terms.

For the real corrections we  use the same observation already discussed in the top case:
the dominant terms suppressed by powers of $T/M$ arise
from configurations where particles have energy $\sim T\ll M$; and kinematics of $2\leftrightarrow 2$ scatterings and $3\leftrightarrow 1$ decays allows only one particle in turn
to have  small energy, such that we only need terms with one power of the thermal distributions in~(\ref{sys:gaugereal}). 
In this way we split the thermal real contributions
in three parts,   $K_0 \sim 0$, $P_H^0\sim 0$ and $P_L^0\sim 0$. 

It turns out that the divergences of the real contributions from the $K_0 \sim 0$ region  cancel those of the vertex contribution,
eq.~(\ref{thermal-vertex}), 
and that the divergences of the real contributions from the $P_H^0 \sim 0$ and $P_L^0\sim 0$ regions cancel divergences from $Z_L$. 

We therefore divide the calculation in these two groups  
to show the cancellation of divergences as soon as possible and to handle simpler expressions.
\begin{itemize}
\item {\it $K_0\sim 0$ region and vertex contributions.} The real contribution from the $K_0\sim 0$ region emerges from (\ref{Gammarealg1}) and (\ref{Gammarealg3})
and corresponds to $z\sim 0$. Here we follow the same technique as in the top case: we expand 
$F_g(y,z)$ in powers of $z$ including only those few powers that can modify the result up to ${\cal O}(T^2/M^2)$. 
We find
\begin{eqnarray}
 & &\Gamma^0 _{\rm tree}\frac{g^2}{8 \pi^2} \left(e^{-\gamma_E/2}\frac{M}{{\bar \mu}}\right)^{d-4} \frac{\Gamma(d/2-2) \Gamma(d-4) }{\Gamma(d-3)} 
\left[8 \left(\frac{2T}{M}\right)^{d-4}\mbox{Li}_{d-4}(1)+\right.\nonumber  \\ 
& &\qquad  \left. +\left(\frac{2T}{M}\right)^{d-2} (d-4) (d-3)^2 (d-2) \mbox{Li}_{d-2}(1)\right], \label{mostro}
 \end{eqnarray}
where $\mbox{Li}_{n}(z)$ is the polylogarithm. Expanding in $d=4-2\varepsilon$ around $d=4$ one gets
a variety of infra-red divergent terms of the  form $1/\varepsilon^2, 1/\varepsilon, (\ln T)/\varepsilon, T^2/\varepsilon, \ln T, \ln^2T$
as well as finite terms of the form $T^2$ and $T^2\ln T$.


A similar approach allows to evaluate the vertex contribution. The real $K_0\sim 0$  and vertex contributions are separately lengthy expressions containing soft and collinear divergences, 
which manifest as poles at $d=4$; however, a spectacular cancellation
takes place and their sum is finite and simple\footnote{Furthermore eq.~(\ref{mostro}) has a pole at $d=5$ because of the singular behavior of Li$_n(1)$ for $n\rightarrow 1$. This emerges at finite
temperature when the purely thermal singularity, $n_B(E)$ for $E\rightarrow 0$, hits those that are also present at $T=0$. This pole, as well as the poles at $d=4$, cancels when adding up with the vertex contribution.}:
\beq \Gamma^0 _{\rm tree} \frac{2 + \ln2}{3}g^2 \frac{T^2}{M^2}.\label{K0V}\eeq

\item
{\it $P_H^0 \sim 0$ and $P_L^0\sim 0$ regions and $Z_L$ contribution.} The real contributions from the $P_H^0 \sim 0$ and  $P_L^0\sim 0$  regions emerge from (\ref{Gammarealg1})-(\ref{Gammarealg2})
 and (\ref{Gammarealg1})-(\ref{Gammarealg4}) respectively and correspond to the points $(y,z)=(1,1)$ and $(y,z)=(0,1)$ respectively. Thus, as in the top case we can now perform a Taylor expansion around
those points. The result is rather complicated and contains a pole at $d=4$ but its sum with the $Z_L$ contribution gives again a simple finite result:
\beq -\Gamma^0 _{\rm tree} \frac{1/2 + \ln2}{3}g^2 \frac{T^2}{M^2}. \label{PHL0Z}\eeq
\end{itemize}
Summing now (\ref{K0V}) and (\ref{PHL0Z}) with the thermal mass contribution we obtain the result in~(\ref{gauge-result}):
\beq \Gamma^0 _{\rm tree} \frac{2 + \ln2}{3}g^2 \frac{T^2}{M^2}-\Gamma^0 _{\rm tree} \frac{1/2 + \ln2}{3}g^2 \frac{T^2}{M^2}-\Gamma^0 _{\rm tree}\frac{1}{2}g^2 \frac{T^2}{M^2}=0.\eeq

\section{Conclusions}\label{concl}
We performed a full NLO computation of one of two key ingredients for thermal leptogenesis: the interaction rate
of the right-handed neutrino.
This unifies partial contributions already included in previous works ($2\to 2$ scatterings, thermal masses)
and gives a very simple result, that we can summarize in one line:
\beq \Gamma_N(E=M) = \label{eq:final}
 \frac{M \lambda^2(M)}{8\pi}
 \bigg[1 +\underbrace{ \frac{29}{32\pi}(3 \alpha_2 + \alpha_Y)}_{3\% (2.5\%)}
 -\underbrace{ \frac{21\lambda_t^2}{32\pi^2}}_{5\%(2\%)}
-  \underbrace{ {\lambda_h}}_{0.1 (0)} \frac{T^2}{M^2}\bigg] 
\eeq  
up to terms suppressed by higher powers of $T/M$ or higher orders. 
At the same order, its  thermal average $\gamma_N$ is given by eq.\eq{gammaGamma}.
This is the only interaction rate that enters in the Boltzmann equations for the right-handed neutrino abundance
and includes all the various decay and scattering terms considered in previous papers.
$\lambda(M)$ is the neutrino Yukawa coupling renormalized at the right-handed neutrino mass $M$.

The numerical values renormalized at the weak scale (at $10^{10}\GeV$) are shown, having 
fixed $\lambda_h = (m_h/2v)^2$ assuming $m_h = 125\GeV$.
The total correction to the final baryon asymmetry ranges between 
$0.5\%$ at $10^{10}\GeV$ and $-3\%$ at the weak scale, slightly depending on the value of $\tilde{m}_1$.
\medskip

The main difference between our result and previous results is the absence of infra-red enhanced corrections of order
$g^2 \ln m_h/M$ (where $m_h \sim gT$ is the thermal higgs mass), which
cancel out as dictated by the KLN theorem
after performing our full quantum and thermal computation at NLO.
In practice this means that  `gauge scatterings' and `higgs scatterings' must be removed from codes for leptogenesis~\cite{GNRRS}:
this makes computations simpler and effectively reduces $\gamma_N$, thus enhancing the efficiency of leptogenesis $\eta \propto 1/\gamma_N$.

Furthermore, $\gamma_N$ crucially enters also in the Boltzmann equation for the lepton asymmetry.
In order to get the whole leptogenesis at NLO a second computation is needed: NLO corrections to the rate of CP-violating processes, which have been addressed at leading order in e.g.~\cite{Liu:1993tg,GNRRS,Nardi:2007jp}. 
This second more complicated computation is not performed in this paper; presumably it can be
performed with the same techniques and the result has the same form, with cancellation of infra-red divergences.
If the quantum version of  Boltzmann equations will turn out to be needed~\cite{quantum}
(which could be possibly the case in the presence of flavor oscillations), 
our precise computation of $\gamma_N$ will remain needed.

%

\paragraph{Acknowledgements} 
This work was supported by the ESF grant MTT8, by SF0690030s09 project, by 
the EU ITN ``Unification in the LHC Era'', contract PITN-GA-2009-237920 (UNILHC) and by MIUR under contract 2006022501.

\footnotesize



\newpage

\centerline{\Large \bf Erratum}

\bigskip\bigskip\bigskip

Our computation in~\cite{v1} has been confirmed by~\cite{Laine:2011pq} up to a difference in one equation:
when computing the reduction in the $N\to LH$ rate due to thermal masses of $L$ and $H$ we incorrectly 
just considered the reduction of its phase space, neglecting that its amplitude is not constant.  The full correct formula had been given in eq.~(89) of~\cite{Giudice:2003jh}:
$$ \Gamma_N(E=M) \stackrel{T\ll M}{\simeq} \frac{M\lambda^2}{8\pi}(1- 2 \frac{m_H^2}{M^2}).$$
This means that below eq.~(18) of~\cite{v1}
$K_T = -( 2m_L^2 + m_H^2 )/ T^2$ gets replaced by $K_T = -2 m_H^2 / T^2$.
After this correction equations (27) and (28) read:
$$
K_T^{\rm higgs} = -1\qquad K_T^{\rm top} = 0
$$
and the coefficient of the thermal effect $T^2/M^2$ in the final result in equation (72) becomes
$$
\Gamma_N(E=M) = \label{eq:final}
 \frac{M \lambda^2(M)}{8\pi}
 \bigg[1 + \frac{29}{32\pi}(3 \alpha_2 + \alpha_Y)
 - \frac{21\lambda_t^2}{32\pi^2}
  -{\lambda_h}\frac{T^2}{M^2}\bigg]
$$
in agreement with \cite{Laine:2011pq}.
These corrections have been fully implemented in the previous text~\cite{v2}.

\end{document}